\documentclass[]{aa}
\usepackage{natbib}
\usepackage{longtable,lscape}
\usepackage{amsmath}
\bibpunct{(}{)}{;}{a}{}{,}
\usepackage{graphicx}
\usepackage{multirow}

\usepackage{graphicx}
\usepackage[colorlinks=true,citecolor=blue]{hyperref}
\usepackage{color}
\usepackage{xspace}
\usepackage{dcolumn}
\usepackage{longtable}
\usepackage{lscape}
\usepackage{soul}
\usepackage{wasysym}
\usepackage{rotating}
\usepackage{afterpage}
\usepackage{mathtools}
\usepackage{blindtext}

\newcommand{\MJup}{M$_{\mathrm{Jup}}$\xspace}

\newcommand{\MSun}{M$_{\odot}$\xspace}

\newcommand{\mic}{$\mu$m\xspace}
\newcommand{\as}{\hbox{$^{\prime\prime}$}\xspace}

\usepackage[varg]{txfonts}
\begin{document}

\title{New spectro-photometric characterization of the substellar object
  HR\,2562\,B using SPHERE\thanks{Based on observations made with European
    Southern Observatory (ESO) telescopes at Paranal Observatory in Chile, under
  programma ID 198.C-0209(D).}}

    \author{D. Mesa\inst{1,2}, J.-L. Baudino\inst{3}, B. Charnay\inst{4}, V. D'Orazi\inst{1}, S. Desidera\inst{1}, A. Boccaletti\inst{4}, R. Gratton\inst{1}, M.Bonnefoy\inst{6}, P. Delorme\inst{6}, M. Langlois\inst{7,8}, A. Vigan\inst{8}, A. Zurlo\inst{9,10,8}, A.-L. Maire\inst{11}, M. Janson\inst{11,12}, J. Antichi\inst{1,13}, A. Baruffolo\inst{1}, P. Bruno\inst{14}, E. Cascone\inst{15}, G.Chauvin\inst{6,16}R.U. Claudi\inst{1},V. De Caprio\inst{15}, D. Fantinel\inst{1}, G. Farisato\inst{1}, M. Feldt\inst{11}, E. Giro\inst{1,17}, J.Hagelberg\inst{6}, S. Incorvaia\inst{18}, E. Lagadec\inst{19}, A.-M. Lagrange\inst{6}, C. Lazzoni\inst{1}, L. Lessio\inst{1}, B. Salasnich\inst{1}, S. Scuderi\inst{14}, E. Sissa\inst{1}, M. Turatto\inst{1}}

    \institute{\inst{1}INAF-Osservatorio Astronomico di Padova, Vicolo dell'Osservatorio 5, Padova, Italy, 35122-I \\
      \inst{2}INCT, Universidad De Atacama, calle Copayapu 485, Copiap\'{o}, Atacama, Chile\\
      \inst{3}Department of Physics, University of Oxford, Oxford, UK \\
      \inst{4}LESIA, Observatoire de Paris, PSL Research University, CNRS, Sorbonne Universit\'{e}s, UPMC Univ. Paris 06, Univ. Paris Diderot, Sorbonne
      Paris Cit\'{e}, 5 Place Jules Janssen, 92195 Meudon, France \\
      \inst{5}Virtual Planetary Laboratory, University of Washington, Seattle, WA 98125, USA \\
      \inst{6}Univ. Grenoble Alpes, IPAG; CNRS, IPAG, 38000 Grenoble, France \\
      \inst{7}Univ. Lyon, Univ. Lyon 1, ENS de Lyon, CNRS, CRAL UMR 5574, 69230 Saint-Genis-Laval, France \\
      \inst{8}Aix-Marseille Universit\'{e}, CNRS, LAM – Laboratoire d’Astrophysique de Marseille, UMR 7326, 13388 Marseille, France \\
      \inst{9}Nucleo de Astronom\'{i}a, Facultad de Ingenier\'{i}a, Universidad Diego Portales, Av. Ejercito 441, Santiago, Chile \\
      \inst{10}Universidad de Chile, Camino el Observatorio, 1515 Santiago, Chile \\
      \inst{11}Max-Planck-Institut f\"ur Astronomie, K\"anigstuhl 17, 69117 Heidelberg, Germany \\
      \inst{12}Department of Astronomy, Stockholm University, SE-10691 Stockholm, Sweden \\
      \inst{13}INAF – Osservatorio Astrofisico di Arcetri, Largo E. Fermi 5, I-50125, Firenze, Italy \\
      \inst{14}INAF – Osservatorio Astrofisico di Catania, Via S. Sofia 78, I-95123, Catania, Italy \\
      \inst{15}INAF – Osservatorio Astronomico di Capodimonte, Via Moiariello 16, I-80131 Napoli, Italy \\
      \inst{16}Unidad Mixta Internacional Franco-Chilena de Astronom\'{i}a, CNRS/INSU UMI 3386 and Departamento de Astronom\'{i}a, Universidad de Chile, Casilla 36-D, Santiago, Chile \\
      \inst{17}INAF Osservatorio Astronomico di Brera, via Emilio Bianchi 46, 23807, Merate (LC), ITALY \\
      \inst{18}INAF – Istituto di Astrofisica Spaziale e Fisica Cosmica di Milano, Via E. Bassini 15, I-20133 Milano, Italy \\
      \inst{19}Universite Cote d’Azur, OCA, CNRS, Lagrange, France \\
}
      
   \date{Received  / accepted }

% \abstract{}{}{}{}{}
% 5 {} token are mandatory
\abstract
   {}
   {HR\,2562 is an F5V star located at $\sim$33~pc from the Sun hosting a
     substellar companion that was discovered using the GPI instrument. The main objective of the present paper is  to
    provide an extensive characterisation of the substellar companion, by deriving its fundamental properties.}
   {We observed HR\,2562 with the near-infrared branch (IFS and IRDIS) of
     SPHERE at the VLT. During our observations IFS was operating in the $YJ$
     band, while IRDIS was observing with the $H$ broad-band filter. The data were
     reduced with the dedicated SPHERE GTO pipeline, which is custom-designed
     for this instrument. On the reduced images, we then applied the post-processing procedures that 
     are specifically prepared to subtract the speckle noise.}
   {The companion is clearly detected in both IRDIS and IFS datasets. We obtained
     photometry in three different spectral bands. The comparison with template spectra
     allowed us to derive a spectral type of T2-T3 for the companion.
     Using both evolutionary and atmospheric models we inferred the
       main physical parameters of the companion obtaining a mass of
       $32\pm14$~\MJup, $T_{eff}$=$1100\pm200$~K and $\log g$=$4.75\pm0.41$.}
   

\titlerunning{Characterizing HR\,2562\,B with SPHERE}
\authorrunning{Mesa et al.}
   \maketitle
%
%________________________________________________________________

\section{Introduction}
\label{intro}

In the last decade, the research field focussed on the 
atmospheric characterization of bound sub-stellar objects (i.e., brown dwarfs and giant planets) 
has experienced an outstanding  boost.
In fact, high-contrast imaging observations have granted the discovery of an 
increasing number of substellar companions \citep{2004A&A...425L..29C,2005A&A...438L..29C,
2008Sci...322.1348M,2010Sci...329...57L,2010ApJ...720L..82B,2013ApJ...779L..26R,
2014ApJ...780L...4B,2015Sci...350...64M,2015ApJ...804...96G,
2017A&A...597L...2M,2017AJ....153...18B} . 
In particular, the new generation
of extreme adaptive optics (XAO) high-contrast imaging facilities such as e.g.,
SPHERE at VLT \citep{2008SPIE.7014E..18B} and GPI at Gemini \citep{2014PNAS..11112661M} 
have proven to be remarkably efficient for this purpose.
Several recent examples of spectroscopic characterization of substellar companions
with those instruments include HD\,95086\,b \citep{2016ApJ...824..121D},
51\,Eri\,b \citep{2015Sci...350...64M,2017A&A...603A..57S},
$\beta$\,Pic\,b \citep{2017AJ....153..182C}, HD\,1160\,B
\citep{2016A&A...587A..56M,2017ApJ...834..162G}, HD\,984\,B
\citep{2017AJ....153..190J}, GJ\,504\,b \citep{2015ESS.....320305B},
GJ\,758\,B \citep{2016A&A...587A..55V}, the four planets of the HR\,8799
system \citep{2016A&A...587A..58B,2016A&A...587A..57Z} and HR\,3549\,B
\citep{2016A&A...593A.119M}. \par
HR\,2562 (HIP\,32775; HD\,50571) is an F5V star with an estimated mass of
M$\sim$1.3 \MSun and a distance of d=33.63$\pm$0.48 pc (see Section~\ref{s:star}). \citet{2006ApJ...644..525M}
identified a debris disk around it by exploiting IRAS and {\it Spitzer} data.
\citet{2014ApJS..211...25C} modeled the stellar spectral energy distribution
(SED) with a two-components disk placed at 1.1 and 341.6~au respectively.
More recently, \citet{2015MNRAS.447..577M}, using {\it Herschel} data, derived a dust radius
of 112.1$\pm$8.4~au with evidence for an inner hole with a radius between
18 and 70~au, and an high inclination of $78.0^{\circ}\pm6.3^{\circ}$. 
\citet{k16} found a substellar companion with mass in the
brown dwarf regime of $30\pm15$~\MJup and effective temperature $T_{\rm eff}$=$1200\pm100$~K. Its
separation of $\sim$0.6\as corresponding to $\sim$20~au from the star put it
into the hole in the disk and its separation and position angle are compatible
with its orbit being coplanar with
the disk itself. \par
HR\,2562 was observed with SPHERE  in order to gather a thorough investigation of the spectral properties 
of its substellar companion, constraining
the physical parameters and fundamental properties, and to obtain new high-precision astrometric data
to better define its orbit and the relation with the disk. In this paper we discuss the spectro-photometric data
obtained for this target, whereas the astrometric data will be discussed in a
second paper (Maire et al., in prep.). \par
The paper is organised as follows: in Sect.\ref{s:star} we discuss
the main physical characteristcs of the host star, 
in Sect.~\ref{s:obs} we describe the observations and data reduction, in
Sect.~\ref{s:res} we report the results that are then discussed in
Sect.~\ref{s:dis}. Finally, in Sect.~\ref{conclusion} we
provide our conclusions.

\section{Host star properties}
\label{s:star}

A careful determination of the fundamental parameters of the host star is crucial
as to obtain reliable and robust estimates  of the substellar companion properties.
In particular, the most crucial issue is the stellar age that
is poorly defined for HR\,2562, as for mid F-type stars in general.
\citet{asiain99} defined an age of 300$\pm$120~Myr classifying HR\,2562 as a member
of the B3 group. A similar age of $\sim$300~Myr was subsequently determined by
\citet{2007ApJ...660.1556R} using space motions, lithium non-detection and
X-ray luminosity. Exploiting the metallicity and the temperature from the
Geneva-Copenaghen survey, \citet{casagrande11} defined a much older age
of 0.9-1.6~Gyr and a similar result of $\sim$0.9~Gyr was found by
\citet{2013A&A...551L...8P} according to the chromospheric activity.
Finally, \citet{moor11} derived an age range of $300_{-180}^{+420}$~Myr using
evolutionary models. 
We tried to determine the main stellar parameters following the
methods described in \citet{desidera15} but more specifically tuned for
a mid-F star as previously done for {HD\,206893\,B in 
  \citet{2017arXiv170900349D}.

\subsection{Spectroscopic parameters}
\label{starspecpar}

In order to carry out a spectroscopic analysis of HR\,2562  
(aiming at deriving radial and rotational velocities, along with atmospheric parameters, $T_{\rm eff}$, log$g$, and metallicity [Fe/H]),
 we retrieved from ESO archive
two UVES spectra\footnote{Prog. ID 096.C-0238(A)}, acquired on the
same night, and two FEROS spectra\footnote{Prog. ID 094.A-9012(A)}, separated by
a few months.
In all cases, data reduction was performed using the on-line pipeline
tools provided by ESO.

From these spectra we have measured the radial velocity (RV) and the projected rotation
velocities (v$sini$), as listed in Table ~\ref{t:rv}, using a custom cross-correlation
function (CCF) procedure.
While our radial and projected rotational velocity estimates agree fairly
well with each other, there is a significant discrepancy (about 6 km/s in RV and
about 25 km/s in v\,sin) with the results obtained with
CORAVEL\footnote{These differences can be explained if the star is an SB2
system with the components seen at similar velocities at the epoch of FEROS
and UVES observations. However, this hypothesis has several difficulties.
There are no significant RV differences between the FEROS and UVES epochs and
the two CORAVEL observations, suggesting a moderately long period but the
period should be short enough to ensure a RV amplitude large enough
to explain the v$sini$ variations. Also, the acquisition images obtained with
SPHERE does not show the presence of equal luminosity companions down to
separation of about 40 mas. Therefore, the projected separation at the epoch
of the SPHERE observations would have been smaller than about 1.35 au. All these
constraints can be satisfied only by a very narrow range of binary parameters.
Furthermore, for an SB2 system with similar components (as required to explain
a variable FWHM of the CCF) an unphysical position on color-magnitude diagram
below main sequence is obtained. We then conclude that additional observations
are needed to settle the issue of the binarity of the central star.}
 \citep{andersen85,demedeiros99,nordstrom04}.

\begin{table*}[!htp]
\caption{Radial velocity and projected rotational velocity}\label{t:rv}
\centering
\begin{tabular}{lccll}
\hline\hline
Date      &  RV (km/s)   & v\,sini (km/s)   & Instrument & Reference \\
\hline
2008-06-14   & 22.90$\pm$0.81 &  60.0      & CORAVEL & \cite{andersen85,nordstrom04}\\
2009-05-29   & 21.58$\pm$0.80 &   60.0     & CORAVEL & \cite{andersen85,nordstrom04} \\
2016-02-23 & 29.28$\pm$0.40  & 31.3   & UVES  & this paper  \\  
2016-03-26 & 27.72$\pm$0.43 & 35.0     & FEROS & this paper  \\ 
\hline
\end{tabular}
\end{table*}

We confirm, in agreement with \citet{asiain99}, that the kinematic parameters
of the star are similar to those of the B3 group when adopting the
\citet{andersen85} RV while they deviate significantly when adopting our own
RV determination.
Concerning the younger groups, no significant membership probability is 
returned using the on-line BANYAN II tool \citep{banyan2} for both
values of system RV.
Taking the uncertainties in system RV into account, we do not consider
the kinematic as constraining the stellar age in the following.

\begin{table}
 \caption{Stellar parameters of HR\,2562.}\label{t:param}
 \centering
\begin{tabular}{lcl}
\hline\hline
Parameter      & Value  & Ref \\
\hline
V (mag)                       &  6.11       & Hipparcos \\
B$-$V (mag)                   &  0.457      & Hipparcos \\
V$-$I (mag)                   &  0.53       & Hipparcos \\
J (mag)                       &  5.305$\pm$0.020  & 2MASS \\
H (mag)                       &  5.128$\pm$0.029  & 2MASS \\
K (mag)                       &  5.020$\pm$0.016  & 2MASS \\
Parallax (mas)                &   29.73$\pm$0.40  & GaiaDR1 \\
$\mu_{\alpha}$ (mas\,yr$^{-1}$)  & 4.872$\pm$0.040  & GaiaDR1 \\
$\mu_{\delta}$ (mas\,yr$^{-1}$)  &  108.568$\pm$0.040  & GaiaDR1 \\
RV  (km\,s$^{-1}$)             &   27.72$\pm$0.43  & this paper \\
$T_{\rm eff}$ (K)      &  6597$\pm$81 & \citet{casagrande11} \\
$\log g$             &  4.3$\pm$0.2 & this paper \\
$\rm [Fe/H]$         &   0.10$\pm$0.06 & this paper \\
EW Li (m\AA)         &   21$\pm$5 & this paper \\
A(Li)                &   2.55$\pm$0.20 & this paper \\
$v \sin i $  (km\,s$^{-1}$)          &   33.2$\pm$2.6 & this paper \\
$\log L_{X}/L_{bol} $  &   -5.21$\pm$0.06 & this paper \\
Age (Myr)            &   200-750 & this paper \\
$M_{star} (M_{\odot})$   &  1.368$\pm$0.018  & this paper \\
$R_{star} (R_{\odot})$   &  1.334$\pm$0.027  & this paper \\
\hline\hline
\end{tabular}
\end{table}

The star was also detected by ROSAT, yielding $\log L_{X}/L_{bol}=-5.21\pm0.06$.
This value is within the locus of Hyades stars (Fig.~\ref{f:param},
panel {\it c}).
The chromospheric emission parameter obtained by \citet{gray06}
$\log R_{HK}=-4.55$ (Fig.~\ref{f:param}, panel {\it b}) is likely
overestimated \citep[see][]{desidera15} and cannot give any constraint on
the stellar parameters.

\begin{figure*}
\begin{center}
\includegraphics[width=16cm]{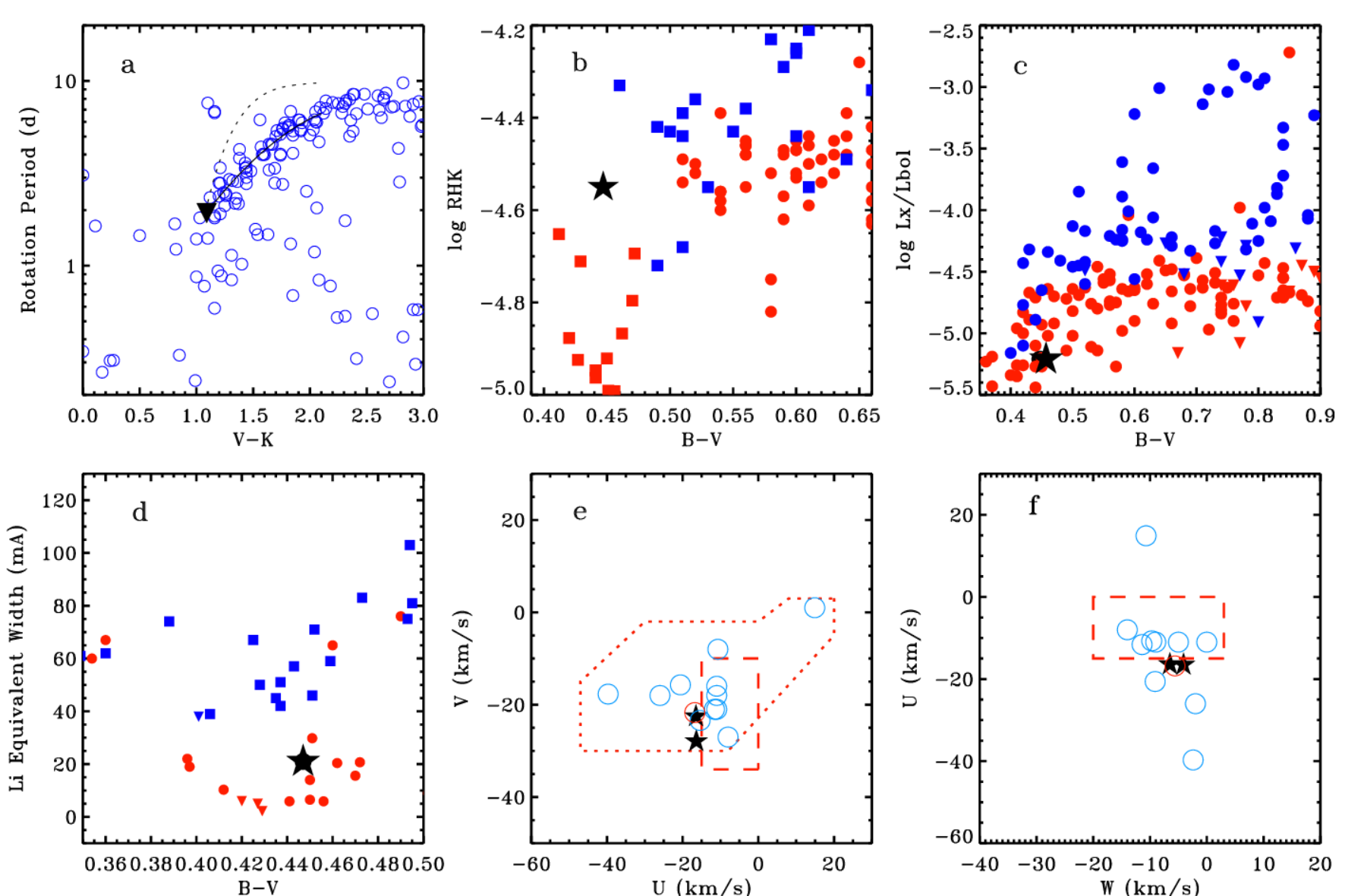}
\caption{Comparison of rotation period, chromospheric emission $\log_{RHK}$, coronal emission
$\log L_{X}/L_{bol}$, Li EW of HR\,2562 to those
of Hyades (red symbols) and Pleiades (blue symbols) open clusters. 
In panel {\it a} (rotation period), the  
triangle represents the upper limits derived from the projected 
rotational velocity. In the same panel the continuous and dashed lines are
the sequences of slow rotators
for Pleiades and Praesepe, respectively \citep{Stauffer16}.
In panels {\it b, c} and {\it d}, showing the chromospheric emission,
$\log L_{X}/L_{bol}$ and lithium equivalent width respectively, circles
and squares represent measurements while upside-down triangles represent upper
limits. The panels {\it e} and {\it f} show 
the kinematic parameters of HR\,2562 for the two values of RV derived in this
paper and from {\it CORAVEL} data, compared to
those of several young moving groups. In these panels, the red open circle
represents the B3 MG while the dotted and dashed lines show the limits of the
young disk as defined by \citet{montes2001} and the young star box as defined
by \citet{zuckerman2004}, respectively.} \label{f:param}
\end{center}
\end{figure*}

The highest-quality FEROS spectrum (median SNR=305 per pixel) was also used to derive the stellar parameters.
The moderately fast rotation makes it challenging to perform a standard
abundance analysis.
Selecting 14 FeI and 2 Fe II isolated lines 
we derived $T_{\rm eff}$=6650$\pm$100~K, $\log g$=4.3$\pm$0.2,
[Fe/H]=+0.13$\pm$0.02 with internal error due to line-by-line scatter, 
and microturbulence $\xi$=1.8 km/s.
The same value of temperature is obtained by fitting the wings of the
$H_{\alpha}$ line, employing the SME code \citep{1996A&AS..118..595V}.
The mild metal overabundance is also supported by Str\"omgren photometry
\citep{casagrande11} with [Fe/H]=+0.07.
The effective temperature derived in that study is 6597$\pm$81~K.
Considering the uncertainties in the spectral analysis for such
a moderately fast-rotator mid F star, we adopt this latter value.
Repeating the abundance analysis for the adopted $T_{\rm eff}$ yields
[Fe/H]=+0.10$\pm$0.02.\par
Isochrone stellar age and stellar mass were derived exploiting
the \citet{bressan12} stellar models and the {\it PARAM} interface
\citep{param} with the input parameters above.
The stellar age is not well constrained (630$\pm$540~Myr), as expected for
a star that lies close to the zero age main sequence (ZAMS).
HR\,2562 is not known as a variable star and Hipparcos photometry shows
a scatter of only 0.007 mag from 118 photometric measurements along the
mission lifetime. Therefore, it is unlikely that variability affects
in significant way the isochrone ages. 

\subsection{Lithium abundance}

The FEROS spectrum was also exploited to measure the lithium content,
a key age indicator for mid-F stars thanks to the presence of the Li $dip$
feature \citep{1986ApJ...302L..49B}. Indeed a marked drop in Lithium
abundance is seen for stars in a narrow range around
$T_{\rm eff}$ 6660 K. The Lithium dip is clearly observed in the Hyades and
other older open clusters \citep{1995ApJ...446..203B}; it starts to be
present at
the age of the open cluster M35 \citep{steinhauer2004}  but it is not seen
in the Pleiades and for younger clusters \citep{boesgaard1988}.
We performed the first measurement of lithium content in HR\,2562, obtaining
EW Li = 21$\pm$5 $m\AA$ and, through spectral synthesis 
as described in \citet{2011A&A...529A..54D,2017A&A...598A..19D},
A(Li) = 2.55$\pm$0.2.
As for HIP\,107412 that has similar $T_{eff}$ (Delorme et al., in press),
lithium provides the tightest constraints to the stellar
age. Comparison with lithium observed in other intermediate age clusters
and associations 
indicates that the Lithium content of HR\,2562 is close to the upper
boundaries of the locus of the Hyades, close to the lower boundaries of that
of M35 \citep{2004ApJ...614L..65S} and well below the locus of the 
Pleiades (see Fig.~\ref{f:param}, panel {\it d}).
While the scatter in the Li-Teff relationship is significant for all the
clusters, preventing a well-defined age calibration, this result confines
the age of the star between 200 and 750 Myr. In the following we will
  consider these age limits in the derivation of the companion properties
  as well as 450~Myr as representative intermediate value.
This age estimate is consistent (but more constraining) with the other 
dating techniques presented above and is slightly narrower than that
adopted by \citet{k16}.
The stellar mass and radius derived as above, altough allowing only this age
range, \cite[see][]{desidera15} are 1.368$\pm$0.018~$M_{\odot}$ and
1.334$\pm$0.027~$R_{\odot}$, respectively.

%__________________________________________________________________
\section{Observations and data reduction}
\label{s:obs}

HR\,2562 was observed with SPHERE on February 6 2017, 
using a non-standard configuration for the IRDIFS mode. In this case, 
IFS \citep{Cl08} was operating in the Y and J bands between 0.95 and 1.35~\mic
while IRDIS \citep{Do08} was using the H broad-band filter configuration
instead of the standard dual band configuration in H2-H3 filters
\citep{2010MNRAS.407...71V}. This choice
was done with the aim to image the disk. The IFS data
were composed by 16 datacubes of 5 frames with an exposure time of 64 s;
those of IRDIS were composed by 16 datacubes of 20 frames with an
exposure time of 16 s. The IRDIS observations were performed using a
$4\times4$ dithering pattern while no dithering was used for IFS. To be able
to use the angular differential imaging \citep[ADI; ][]{2006ApJ...641..556M}
technique, the field of view (FOV) was allowed to rotate during the
observations, with the pupil fixed with respect to the detector.
To maximize the total rotation of the FOV, we observed the star
during its passage to the meridian with a total rotation of$\sim34.9^{\circ}$.
For both IFS and IRDIS we acquired observing frames with the star image
off-centered with respect to the coronagraph, to be able to calibrate the
companion flux. In order to avoid saturation, 
a neutral density filter was employed.
Moreover, we took frames with four satellite spots that are symmetrically
located with respect to the central star by a suitable modification of the
wavefront provided by the AO. This is done to properly determine the center of
each frame: the use of satellite spots  was first proposed by
\citet{2006ApJ...647..620S} and \citet{2006ApJ...647..612M} We refer to 
\citealt{2013aoel.confE..63L} and 
\citealt{mesa2015} for an extended discussion of their use within the SPHERE framework. \par
Both IFS and IRDIS data were reduced using the SPHERE data center
\footnote{\url{http://sphere.osug.fr/spip.php?rubrique16&lang=en}}
applying the appropriate calibrations following the SPHERE data reduction and
handling \citep[DRH,][]{2008SPIE.7019E..39P} pipeline. More specifically, 
IRDIS dataset requires the application of dark and flat-field frames and the definition of
the star center. On the other hand, for IFS observations the reduction steps include: dark and flat-filed correction, 
definition of the spectral positions, wavelength calibration, and the
application of the instrumental flat. For IFS we then apply on the wavelength-
calibrated datacubes (each of them composed by 39 monochromatic frames) the
speckle subtraction exploiting the principal components analysis
\citep[PCA,][]{2012ApJ...755L..28S} as described in \citet{mesa2015} and
\citet{zurlo2014}. For what concerns IRDIS we performed the speckle subtraction
with both PCA and TLOCI \citep{2014SPIE.9148E..0UM}, using the
consortium pipeline (Galicher et al., in prep.).
%________________________________________________________________________

\section{Results}
\label{s:res}

The data reduction procedure described in Sect.~\ref{s:obs} provided
the IFS and IRDIS final images displayed in Fig.~\ref{f:finalim}. The
companion is clearly detected with both instruments. However, it
was not possible to image the debris disk, despite the fact we decided to employ the IRDIS H broad-band filter for this specific purpose.  
The companion was detected
with a S/N of the order of 30 with IFS, and with a S/N of 20 with
IRDIS. In both cases, we have exploited the PCA algorithm that allowed us to
obtain the best results with respect to other algorithms. 

\begin{figure*}
\centering
\includegraphics[width=0.45\textwidth]{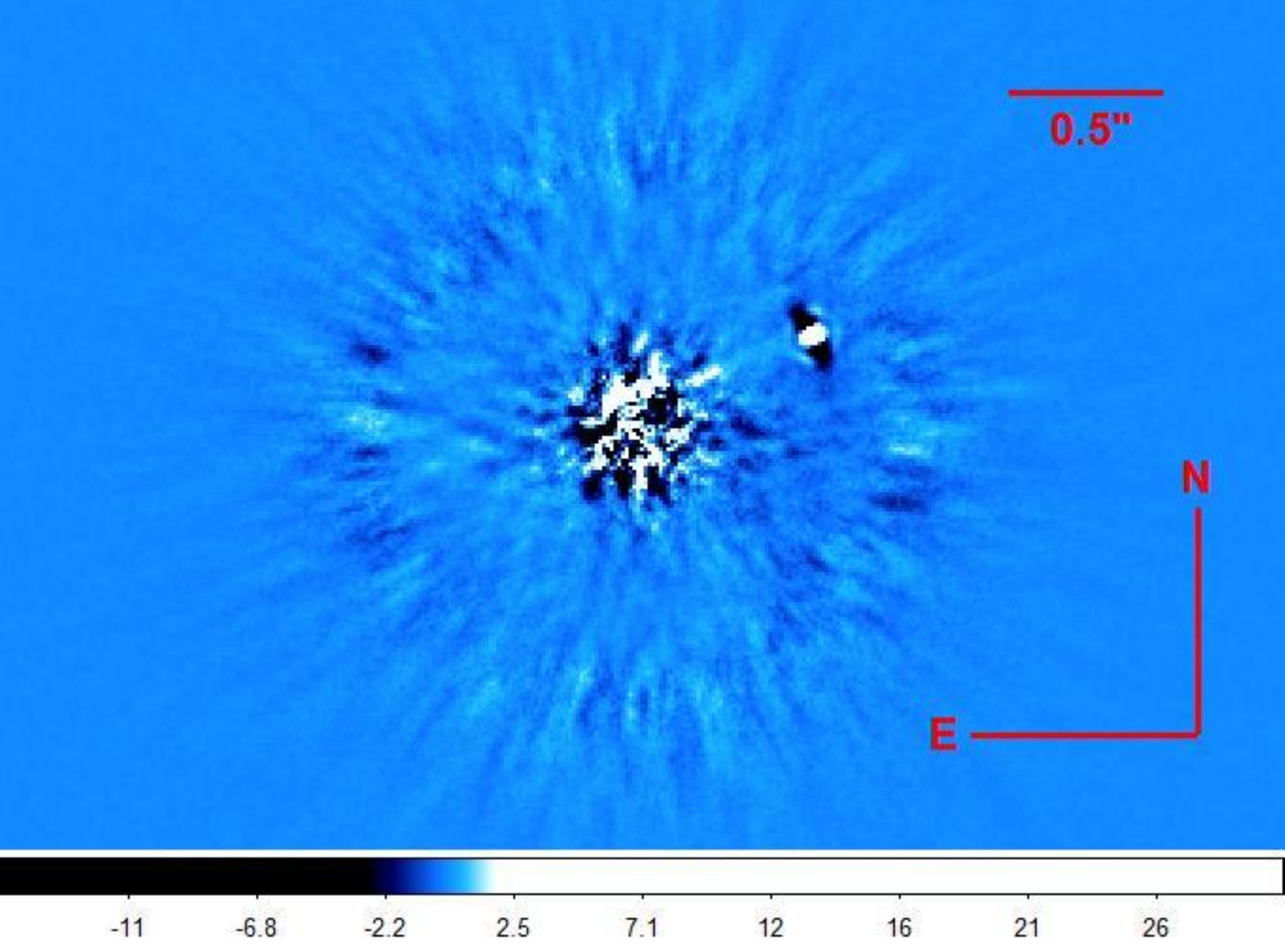}
\includegraphics[width=0.45\textwidth]{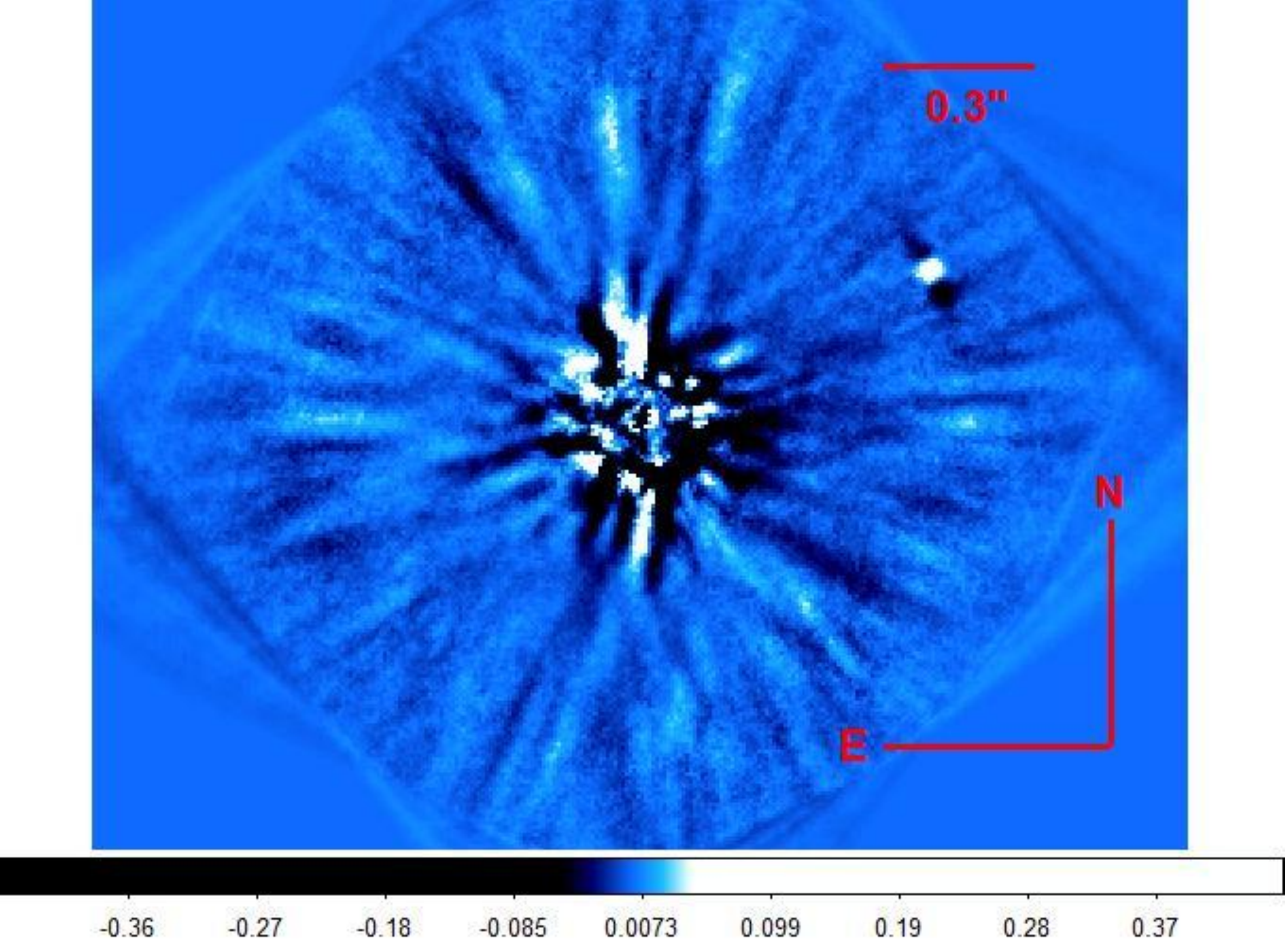}
\caption{{\it Left:} Final image obtained with IRDIS.{\it Right:} Final image obtained with IFS. Both images are obtained using the PCA algorithm.}
\label{f:finalim}
\end{figure*}

\begin{figure}
\centering
\includegraphics[width=0.45\textwidth]{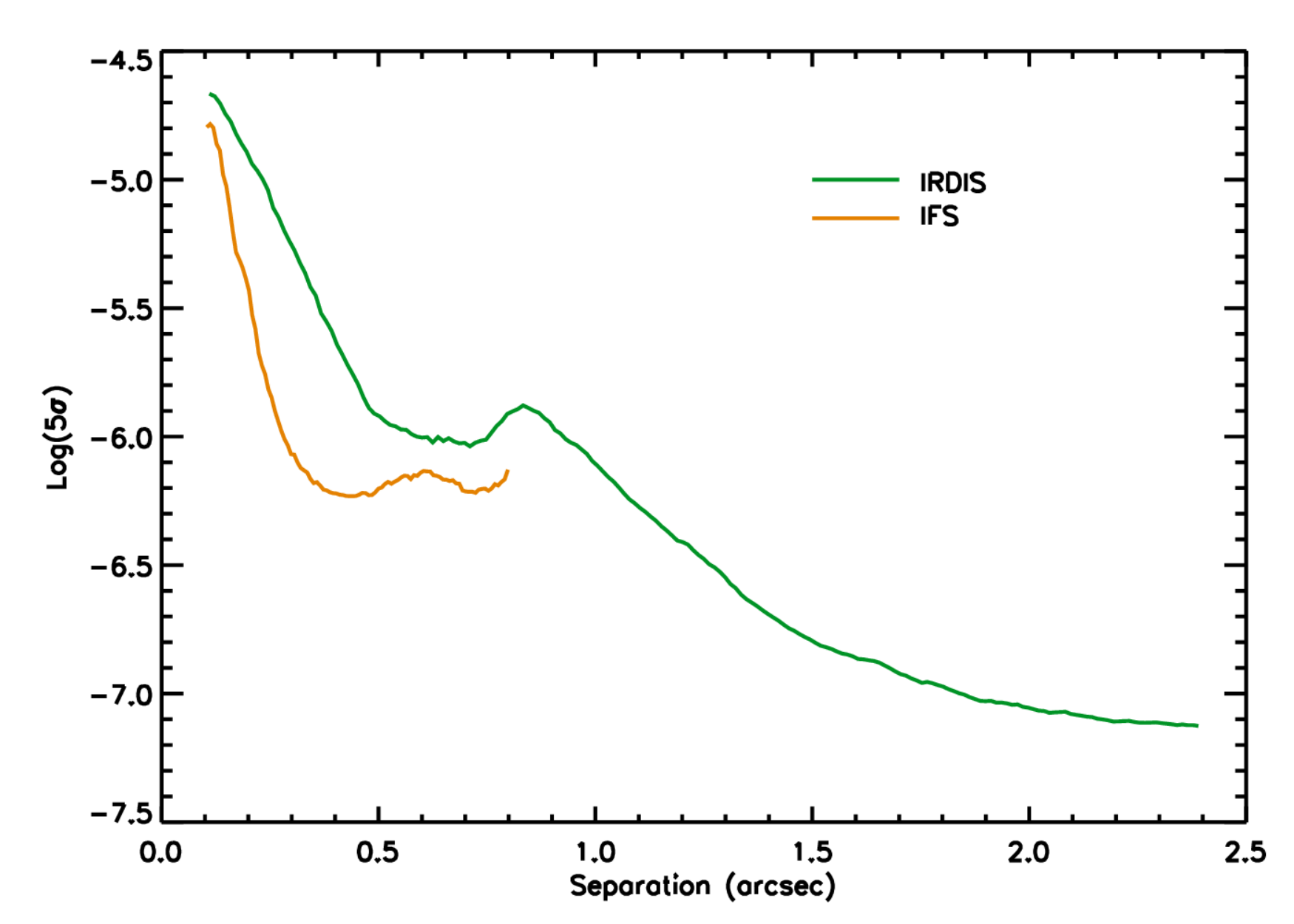}
\caption{Contrast plot for IRDIS (green line) and IFS (orange line).}
\label{f:contrast}
\end{figure}

Moreover,  we were able to define from the final images the contrast limit
for IFS and IRDIS following the procedure devised in \citet{mesa2015}.
The contrast values obtained was corrected to take into account the
self-subtraction of the high-contrast imaging method. This was evaluated
by injecting in the original datacube simulated planets of known flux.
The resulting limiting contrast curves are displayed in Fig.~\ref{f:contrast}
where we show that we could obtain a contrast better than $10^{-6}$ at a
separation of $\sim$0.3\as or larger with IFS. As for IRDIS, we reach 
a contrast better than $10^{-7}$ at separation larger than 1.5\as. \par
For the companion, we obtained photometric measurements for each spectral channel
introducing a negative simulated planet in the original dataset at the
companion position and running our PCA procedure. As a cross-check, we performed
the same procedure using a classical ADI procedure and calculated the aperture
photometry of the companion on the final images, after taking properly into
account the self-subtraction. The three methods provide similar results and
we list the final photometry obtained using the first method
described above, for the Y, J and H bands (here and in all the
following analysis we are using the 2MASS photometric system)
in Tab.~\ref{t:photo}.
The photometric errors on these values are mainly due to uncertainties on the
attenuation factor of the method, to variations on the stellar PSF and of the
stellar speckles noise during the observing sequence.

\begin{table}
  \caption{Absolute magnitude for HR\,2562\,B in Y, J and H band. }
  \label{t:photo}
  \centering
\begin{tabular}{c c c}
  \hline
  \hline
$M_{Y}$   &   $M_{J}$   &   $M_{H}$   \\
\hline
15.97$\pm$0.33  &  15.01$\pm$0.11  &   13.98$\pm$0.14   \\           
\hline
\end{tabular}
\end{table}

From the photometric data, we obtained a low resolution spectrum for the
companion that was then converted from contrast to flux by multiplying it
by a flux-calibrated BT-NEXTGEN \citep{2012RSPTA.370.2765A} synthetic spectrum
for the host star, adopting $T_{eff}$=6400~K, $\log{g}$=4.0 and [M/H]=0.0 that
gives the best fit with the SED of the star.
The final result of this procedure is displayed in Fig.~\ref{f:spectrum} and
compared with the GPI
spectrum from \citet{k16}. In the J-band, the SPHERE IFS and GPI spectra are
very similar even if the peak at $\sim$1.27\mic is slightly higher
($\sim$15\% but however within the error bars - see Fig.~\ref{f:spectrum}) in
the SPHERE IFS data. In the H band, in order to compare the IRDIS broad band
result with GPI data, we integrated the GPI spectrum over the wavelength range
of the IRDIS H broad band filter. In this case the IRDIS value is about 35\%
higher than that from GPI. \par
It is noteworthy, however, that systematic errors could be present
between GPI and SPHERE probably due to differences in the algorithms for the
spectral extraction and/or in the flux normalization procedure, as
highlighted by \citet{2017AJ....154...10R} who compared the GPI 
\citep{2015Sci...350...64M} and SPHERE \citep{2017A&A...603A..57S} results for
51\,Eri\,b. For this reason, in the following analysis we will be using only the
SPHERE data.

\begin{figure}
\centering
\includegraphics[width=0.45\textwidth]{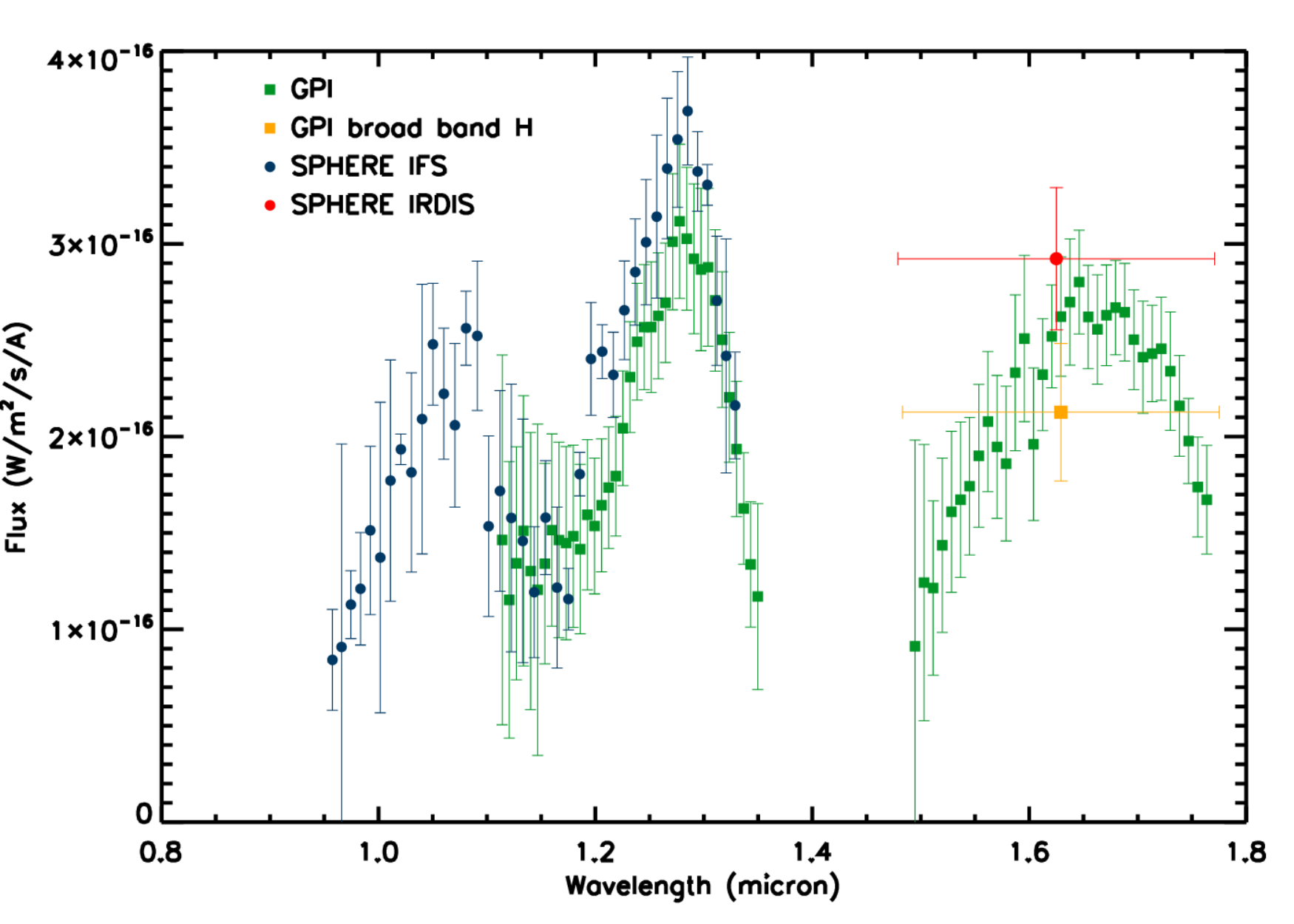}
\caption{Low resolution spectrum for HR\,2562\,B (blue circles) compared to
  the GPI spectrum for the same object (green squares). The red circle is the
  spectral point derived from IRDIS where the large error band on the
  wavelength scale denotes
  that it is derived with a broad band filter. The orange square is derived from
  the GPI data considering the same wavelength range of the IRDIS broad band
  filter to be able to make a comparison between the two instruments.}
\label{f:spectrum}
\end{figure}

%_____________________________________________________________________

\section{Discussion}
\label{s:dis}

\subsection{Characterization of HR\,2562\,B}
\label{s:companion}
\subsubsection{Color-magnitude diagram}
\label{s:colmagdiagram}
Starting from the photometric values listed in Tab.~\ref{t:photo}, we have
produced the color-magnitude diagram ($J_s$ $vs.$ $J_s$-H). The use
of the $J_s$ magnitude is justified in this case by the fact that it
allows to better discriminate objects near the L/T transition from background
contaminants.
 In this diagram, the position of
HR\,2562\,B is compared to those of M, L and T field dwarfs, whose photometry
has been gathered from the flux-calibrated near-infrared spectra from
the SpeXPrism library \citep{2014ASInC..11....7B}. Like e.g. in 
\citet{2016A&A...587A..57Z}, we smoothed these spectra to the IRDIS resolution
using the IRDIS passbands, a model of the Paranal atmospheric transmission
using the ESO Skycalc web application
\footnote{\url{ http://www.eso.org/observing/etc/bin/gen/form?INS.MODE=swspectr+INS.NAME=SKYCALC}}\citep{2012A&A...543A..92N,2013A&A...560A..91J}
and a model spectrum of Vega \citep{2007ASPC..364..315B}.
We also show the positions
of substellar companions with an age comparable to that of HR\,2562\,B
\citep{2007ApJ...654..570L,2010MNRAS.405.1140G,2011ApJ...729..139W,
  2013ApJ...774...55B,2014ApJ...787....5N,2015ApJ...804...96G,
  2015MNRAS.454.4476S,2017AJ....153...18B} as stellar age is known to be
a relevant parameter for the spectral and photometric characterization of
substellar objects. HR\,2562\,B
sits at the L-T transition, very near to the location of (although slightly
redder than) HN\,Peg\,B that has comparable age 
(300-400~Myr) and mass ($\sim$20\MJup), according to \citet{2007ApJ...654..570L}. 

\begin{figure}
  \begin{center}
\centering
\includegraphics[width=\columnwidth]{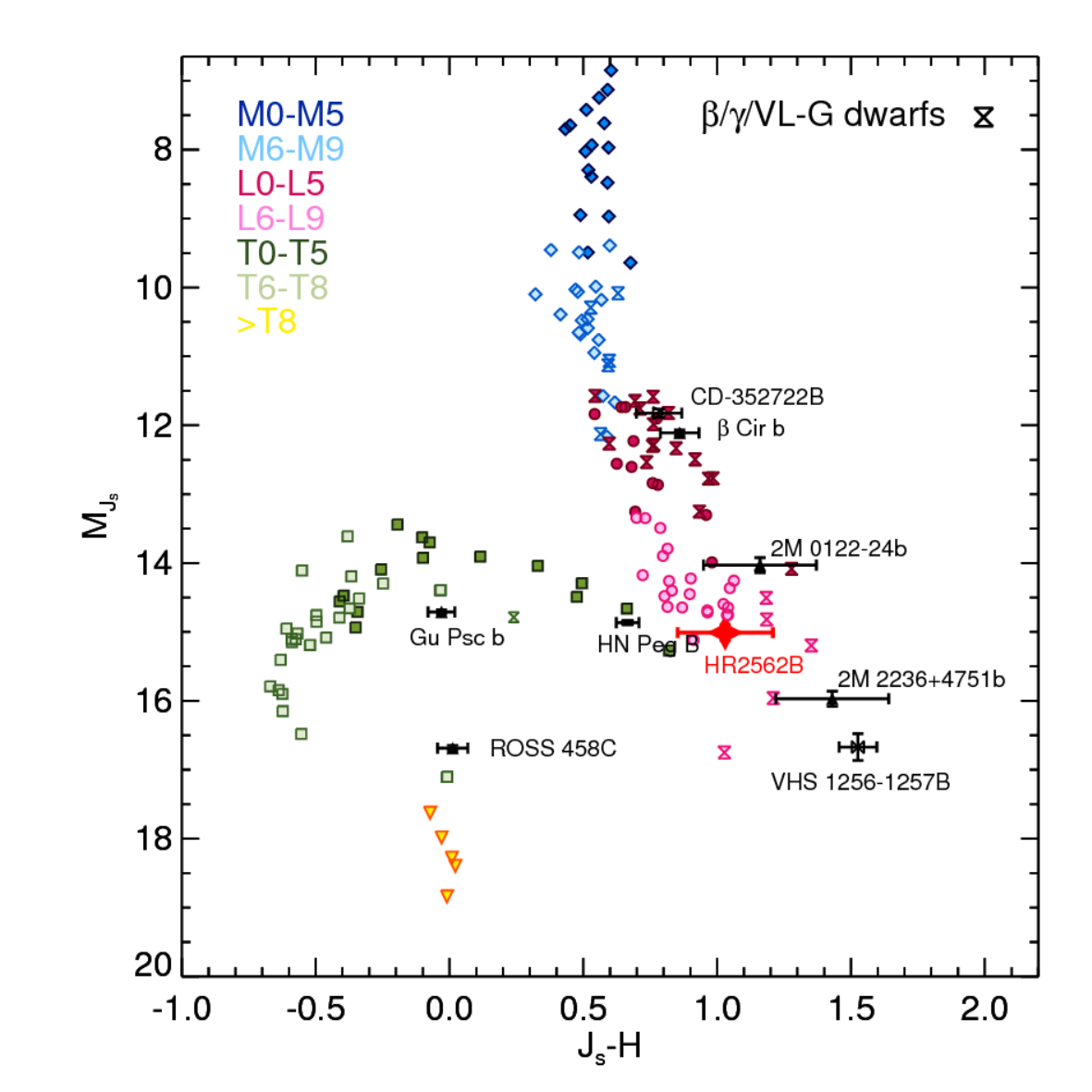}
\caption{$J_s$ versus $J_s$-H diagram showing the position of HR\,2562\,B,
  represented by a red star, with
  respect to that of M, L and T field dwarfs and of young known companions. The
  black asterisks show the positions of substellar companions with an age
  comparable to that of HR\,2562\,B that is indicated by the red point.}
\label{f:colmag}
\end{center}
\end{figure}

\subsubsection{Companion parameters through evolutionary models}
\label{s:atmomodel}

We estimated the main
parameters of the companion using the AMES-COND models
\citep{2003IAUS..211..325A} and the AMES-DUSTY models
\citep{2001ApJ...556..357A}, starting from photometry reported in Tab.~\ref{t:photo} and
assuming the age we have discussed in
Sec.~\ref{s:star}.  The results for the companion mass and its
surface gravity are listed in Tab.~\ref{t:mass} and in Tab.~\ref{t:massdusty}.
For these calculations, we adopted three different ages that cover
the whole age range proposed in Sec.~\ref{s:star} for HR\,2562.
The Y and J band results are in good agreement with each
other, while the H band determination tends to provide different masses
especially in the case of the AMES-COND models.
For the intermediate age of 450~Myr the AMES-COND model provides a value
of the mass between 20 and 30~\MJup ($\log g$$\sim$4.7 dex, with $g$ in cgs), which is
slightly lower than that obtained by \citet{k16} because of the younger
age adopted in this paper. On the other hand, the AMES-DUSTY model provides
a larger value of 40-46~\MJup ($\log g\sim$5.0 dex).
These values for the mass are in good agreement with what was recently
found from \citet{2017ApJS..231...15D} that, evaluating the dynamical masses
of ultracool dwarfes, found for BDs with spectral type around T2-T3 (see
Section~\ref{spectraltemplate}) a mass range around 30-40~\MJup.

\begin{table}
  \caption{Estimation of the companion mass(in \MJup) and log(g) using the
    Y, J and H band photometry and the AMES-COND models.}
  \label{t:mass}
  \centering
\begin{tabular}{c c c c c c c}
  \hline
  \hline
Age  &  \multicolumn{3}{c}{Mass (\MJup)} & \multicolumn{3}{c}{log(g)} \\
(Myr)&  Y   &   J   &   H  &  Y   &   J   &   H  \\
\hline
200  &   11.46  &  11.80  &  18.55 & 4.34  &  4.35  &  4.58   \\
450  &   20.85  &  22.12  &  29.16 & 4.69  &  4.71  &  4.85   \\
750  &   27.40  &  28.67  &  37.58 & 4.85  &  4.87  &  5.01   \\
\hline
\end{tabular}
\end{table}

\begin{table}
  \caption{Estimation of the companion mass (in \MJup) and log(g) using the
    Y, J and H band photometry and the AMES-DUSTY models.}
  \label{t:massdusty}
  \centering
\begin{tabular}{c c c c c c c}
  \hline
  \hline
  Age   &  \multicolumn{3}{c}{Mass (\MJup)} & \multicolumn{3}{c}{log(g)} \\
  (Myr) &  Y   &   J   &   H  &  Y   &   J   &   H   \\
\hline
200  &   34.04  &  32.14  &  28.41 & 4.82  &  4.80  &  4.74  \\
450  &   46.59  &  44.84  &  40.21 & 5.03  &  5.02  &  4.97  \\
750  &   56.08  &  54.34  &  49.51 & 5.18  &  5.16  &  5.12  \\
\hline
\end{tabular}
\end{table}

\subsubsection{Fitting with spectral template}
\label{spectraltemplate}

In order to better ascertain the spectral type of HR\,2562\,B, we have 
performed fit procedures of its
spectrum with sample spectra
of field BDs taken from the {\it Spex Prism spectral Libraries}\footnote{\url{http://pono.ucsd.edu/~adam/browndwarfs/spexprism/}}
\citep{2014ASInC..11....7B}. The final result is displayed
in Fig.~\ref{f:spectrafit}, where we show the comparison of the extracted
spectrum for HR\,2562\,B with those of ten objects having
spectral types between L5 and T5
\citep{2004AJ....127.2856B,2006ApJ...637.1067B,2007ApJ...654..570L,
  2008ApJ...678.1372C,2009ApJ...702..154S,2013ApJ...777L..20L,
  2014MNRAS.445.3694D,2014ApJ...787....5N,2015ApJ...808L..20G}.
The best fit was obtained
with the spectrum of HN\,Peg\,B \citep{2007ApJ...654..570L} that is classified
as T2.5 spectral type. This result confirms what we obtained in
Sec.~\ref{s:colmagdiagram} where we showed (see Fig.~\ref{f:colmag}) that the
positions of these two objects are very similar in the color-magnitude diagram.
Thus, we found compelling evidence that they could be very similar objects. However, a
comparably good fit was also obtained with the spectrum of
SDSS\,J143553.25+112948.6 \citep{2006AJ....131.2722C} that was classified as
T2$\pm$1 spectral type, and SDSS\,J102109.69030420.1AB
\citep{2006ApJ...637.1067B} that was classified as a T3 spectral type. From
these results we can define this object as a T2-T3 spectral type. It is noteworthy, however, 
that the value derived from the IRDIS H broad band data
does not agree well with this result. Indeed, looking at the comparison of
the flux in H broad band between HR\,2562\,B and the template spectra
represented by the dark and the orange squares in Fig.~\ref{f:spectrafit}
respectively, our finding seems in better agreement with a late-L spectral type.

\begin{figure}
  \begin{center}
\centering
\includegraphics[width=0.9\columnwidth]{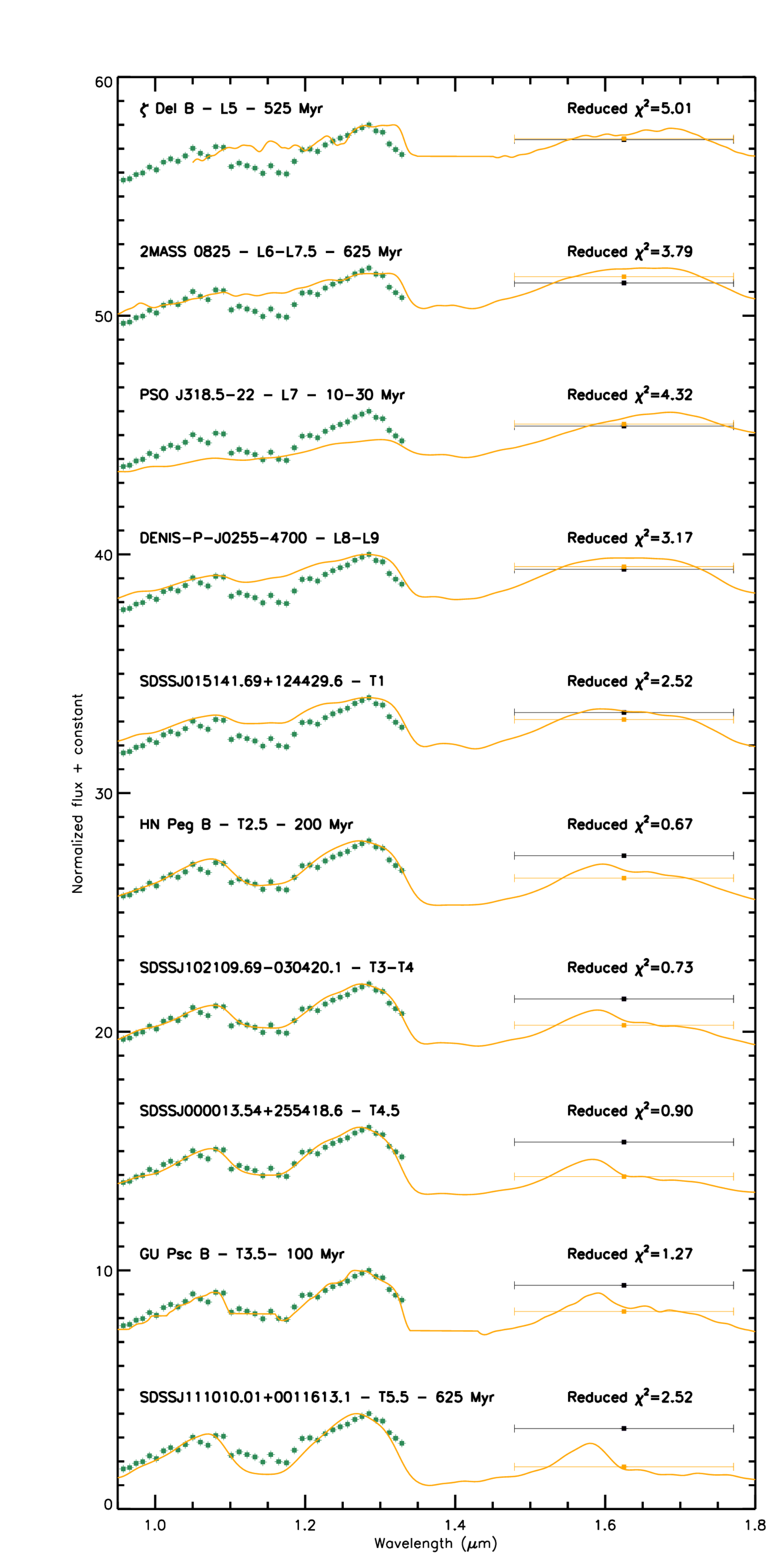}
\caption{Comparison of the extracted spectrum of HR\,2562\,B (green squares)
  to those of 10 template spectra (orange lines) with spectral types ranging
  between L5 and T5.5. The dark square represents the value of the IRDIS
  H broad band photometry while the orange square is the H broad band photometry
  for the compared objects calculated starting from the original spectrum.}
\label{f:spectrafit}
\end{center}
\end{figure}

\subsubsection{Fitting with atmospheric models}
\label{s:synt}

To further constrain the physical parameters of HR\,2562\,B we compared its
extracted spectrum to synthetic spectra from different atmospheric models. \par
We first used the BT-Settl models \citep{2014IAUS..299..271A} with a grid
covering $T_{eff}$ between 900 and 2500~K with a step of 100~K and a $\log g$
ranging between 2.5 and 5.5 dex, with a step of 0.5. All the models were for
a solar metallicity. The results of this procedure
are displayed in Fig.~\ref{f:spectrafitsynt}. The
best fit is with the spectrum with $T_{eff}$=1200~K and $\log{g}$=4.5.
A good fit is however obtained even
for a spectrum with $T_{eff}$=1100~K and $\log{g}$=4.0 and for a spectrum
with $T_{eff}$=1300~K and $\log{g}$=5.5. We can then conclude that the fit
gives $T_{eff}$=1200$\pm$100~K and $\log{g}$=$4.5_{-0.5}^{1.0}$. $T_{eff}$
and log(g) values found through this procedure are in good agreement with those
found by \citet{k16}.

\begin{figure*}
  \begin{center}
\centering
\includegraphics[width=0.9\textwidth]{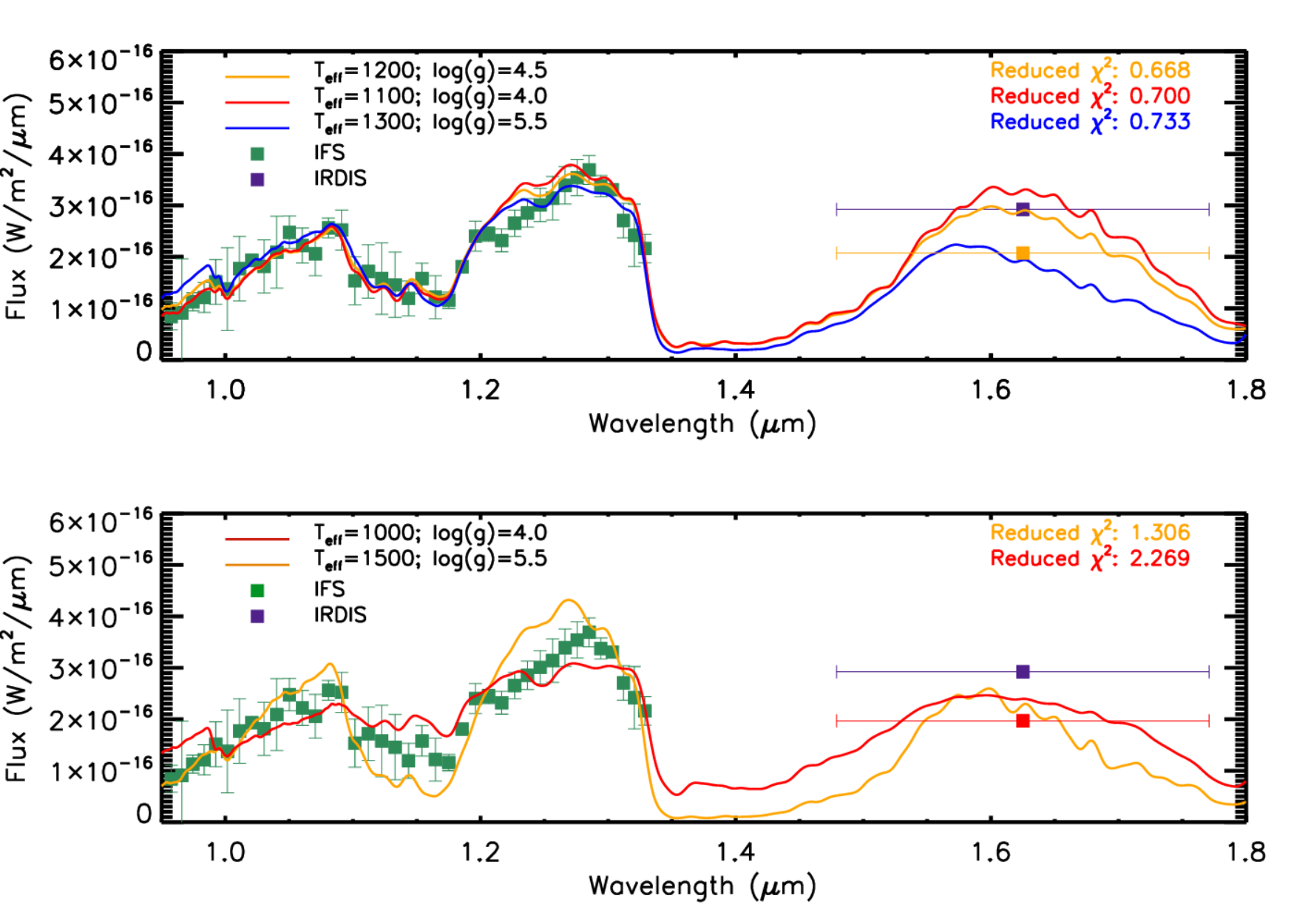}
\caption{{\it Upper panel: } Spectrum extracted for HR\,2562\,B
  (green squares)
  with its error bars. The colored solid lines represents the three best
  fitting spectra from the BT-Settl synthetic spectra. Orange square is
  obtained from the corresponding spectrum calculating the flux on the IRDIS H
  broad band filter to be able to better compare it with the IRDIS data.
  The same operation was not done for the other spectra to avoid confusion
  in the final plot.
  {\it Lower panel: }
  Same as upper panel, but using synthetic spectra with poorer fits.}
\label{f:spectrafitsynt}
\end{center}
\end{figure*}

As a second step, we have compared the extracted spectrum for HR\,2562\,B with the
Exo-REM model \citep{2015A&A...582A..83B}, which is specifically developed for young
giant exoplanets. We first generated a grid of models with $T_{eff}$ between
400 and 1800~K with a step of 100~K, $\log g$ in the range 2.5-5.5 dex with a step
of 0.1, with and without Fe and $Mg_2SiO_4$ considering a particle radius of
30~\mic and assuming equilibrium and non-equilibrium chemistry
($k_{zz}=10^8$~$cm^2s^{-1}$). All these models were then compared to the spectrum of
HR\,2562\,B to find the one that minimize the $\chi^2$. All the
values with $\chi^2$ less than 1 were then considered as acceptable. The best
fits obtained from this procedure are shown in the upper panel of
Fig.~\ref{f:exoremjlb}, while in the lower
panel we show as a comparison an example of a poor quality fit. 
The best results are
for models with $T_{eff}$ in the range 1100-1200~K, $\log g\sim$5.0 dex, cloud with
an optical depth of $\tau_{ref}$=0.5 and non-equilibrium chemistry. However,
considering all the models with a value of $\chi^2$ less than 1 we can only
constrain a $T_{eff}$ in the range 950-1200~K and $\log g$ in the range 3.4-5.2 dex 

\begin{figure*}
\centering
\includegraphics[width=\textwidth]{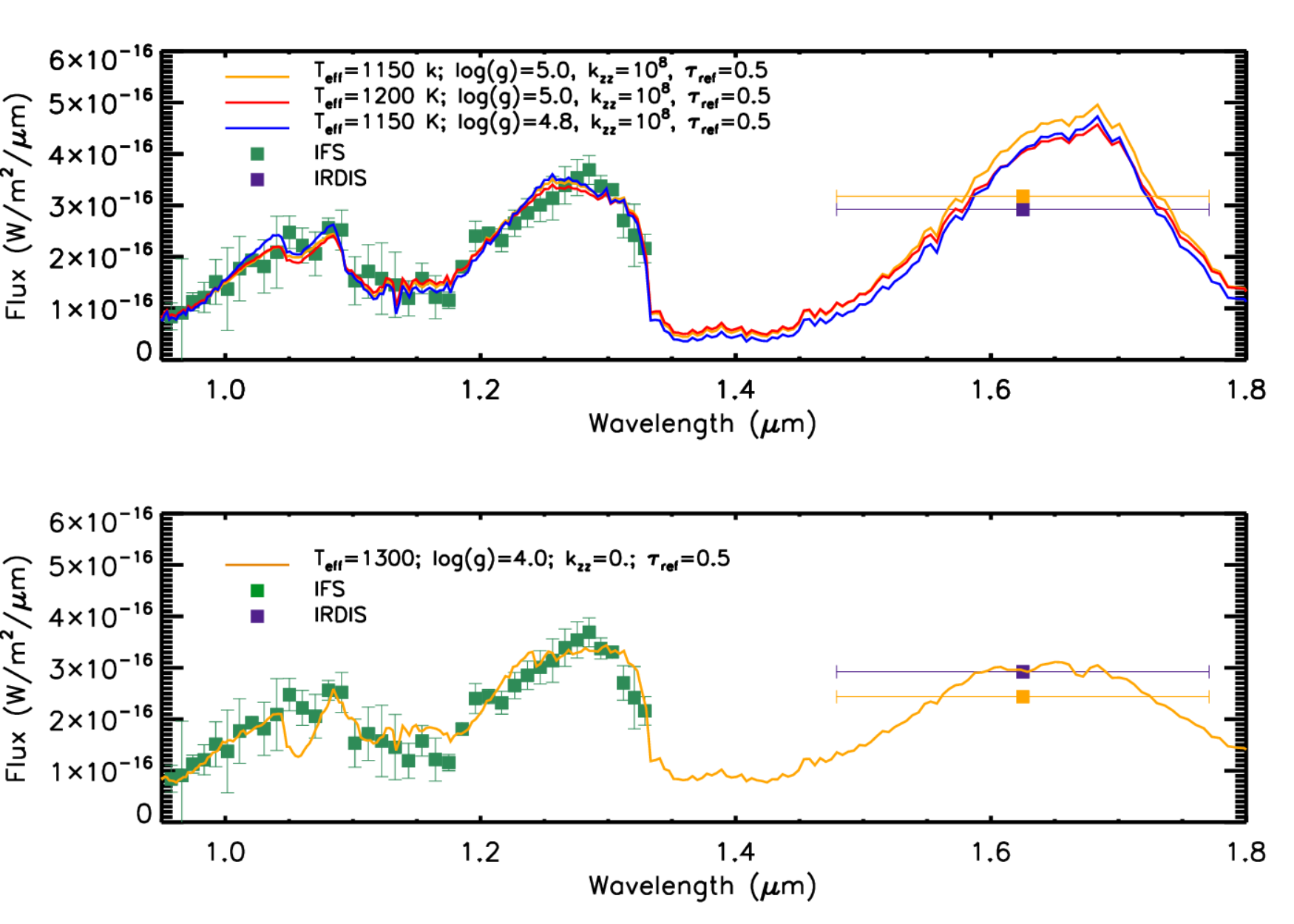}
\caption{Same as Fig.~\ref{f:spectrafitsynt} but for the case of the Exo-REM
  models. The poorer fit in the lower panel is the best case taking into account
equilibrium chemistry.}
\label{f:exoremjlb}
\end{figure*}

Aiming at carefully evaluate the contribution of the cloud coverage on the definition
of the main parameters for the companion, we have also used 
a new version of the Exo-REM models. This includes a self-consistent
modelling of the clouds (both iron and
silicate) that is based on the parameterization of the $k_{zz}$ profile, and considers
cloud condensation, coalescence, sedimentation and vertical mixing
(Charnay et al., in prep.). The size of the cloud particles is calculated
by comparing typical timescales for condensation, coalescence, sedimentation
and vertical mixing, following the method described in \citet{1978Icar...36....1R}.
Moreover, these models take into account cloud scattering and
can also simulate a cloud coverage fraction as done by
\citet{2010ApJ...723L.117M}. The model $T_{eff}$ ranges from 400 to 2000~K
with step of 100~K, while $\log g$ varies from 3 to 5 dex, with step of 0.1 dex. 
We considered models with 0.3, 1 and 3 times the solar metallicity. The
decision to explore different metallicities is due to the fact that
we have found that the star is slightly metal rich with respect to the
Sun (see Section \ref{starspecpar}) and that there are claims in literature
about metallicity enhancement of substellar companions with respect to
their parent stars \citep{skemer2016,2017A&A...603A..57S}. Equilibrium
and non-equilibrium chemistry were adopted along with different cloud
coverage: the best fitting results in models with
non-equilibrium chemistry and a cloud coverage of 95\%. The best fit
spectra from this procedure are shown in Fig.~\ref{f:exorembch} while
the $\chi^2$ map is displayed in Fig.~\ref{f:parspace} together with
  those from BT-Settl and not modified Exo-REM models. The result
of this second procedure points toward an object at $T_{eff}$ between 1000
and 1200~K and $\log g$ just above 4.0 dex. 
Indeed, considering a metallicity
three times the solar value, we obtain $T_{eff}$=1100$\pm$100~K and
$\log g$=4.3$\pm$0.6 dex, while for a solar metallicity we estimate from the
model $T_{eff}$=900-1150~K and $\log g$=4.1$\pm$0.6 dex. For what
concerns the exploration of the metallicity, instead, no definitive
results can be deduced from these models even if models with metallicity
lower than solar tend to prefer values of $\log g$ less than 4 while
models with metallicity higher than solar tend to prefer values of $\log g$
nearer to the values deduced through evolutionary models.\par

\subsubsection{Comparison of results from evolutionary and atmospheric models}
\label{s:comp}

As for $\log g$, by comparing our data with atmospheric models  
we have inferred values lower than that obtained from the evolutionary models, in particular
when DUSTY models are employed (see Section~\ref{s:atmomodel}).
This can be explained by the fact that the cloud coverage of the
atmosphere of HR\,2562\,B is likely only partial, as confirmed by the quite compelling
evidence provided by the Exo-REM models. However, when we consider the whole
atmospheric model data-sets we obtain 
a wide range of $\log g$ ranging from 3.4 to 5.2. We conclude that it is
difficult to firmly constrain the value of $\log g$ by exploiting atmospheric models, because 
their dependence on gravity is relatively small. From this point
of view, the estimates that we gathered using the evolutionary models are much
more useful, and with significantly smaller dispersion. In fact, they
represents two extreme conditions (absence of clouds
in the COND models $vs$ complete coverage of clouds for the DUSTY case)
that can be used to constrain the correct value of $\log g$. We can then
  consider the lower and the upper values obtained from these models and
  listed in Tab.~\ref{t:mass} and \ref{t:massdusty} and use them to
  estimate the value of $\log g$ as $4.76\pm0.42$. The main contribution to the
  error bar is clearly given by the uncertainties on the age of the system.
On the other hand, $T_{eff}$ is much
better constrained with values in the range 900-1300~K so that we can then
assume for $T_{eff}$ a value of $1100\pm200$~K. This is in good agreement
with the spectral classification of T2-T3 that we obtained from the
procedure described in Section~\ref{spectraltemplate}. Indeed,
\citet{2013ApJS..208....9P} foresee $T_{eff}$=1200~K for a T2 spectral type
and of 1160~K for a T3 spectral type.

\begin{figure*}
\centering
\includegraphics[width=\textwidth]{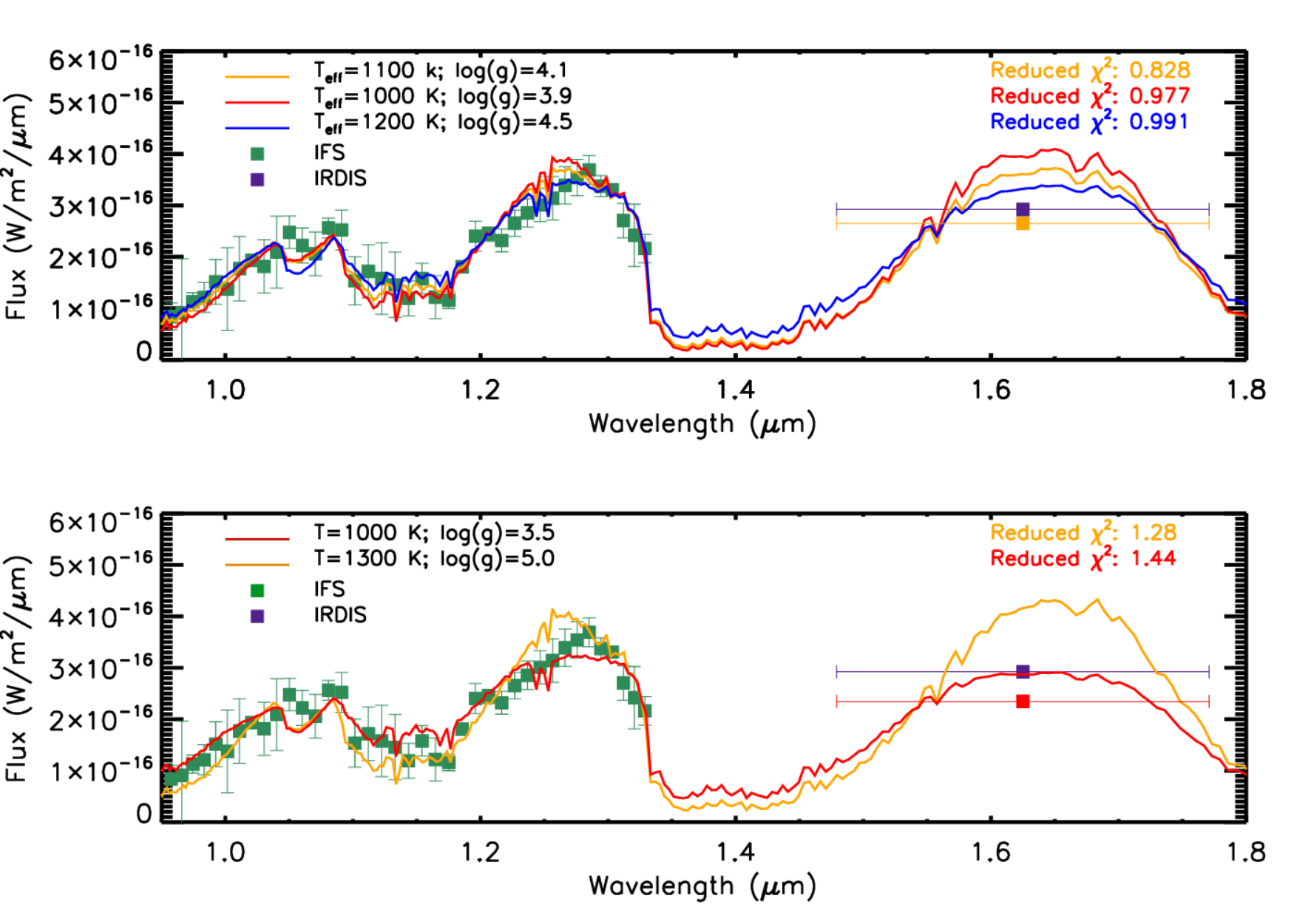}
\caption{Same as Fig.~\ref{f:spectrafitsynt} but for the case of the Exo-REM
models with modified clouds treatment.}
\label{f:exorembch}
\end{figure*}

\begin{figure}
  \centering
  \includegraphics[width=\columnwidth]{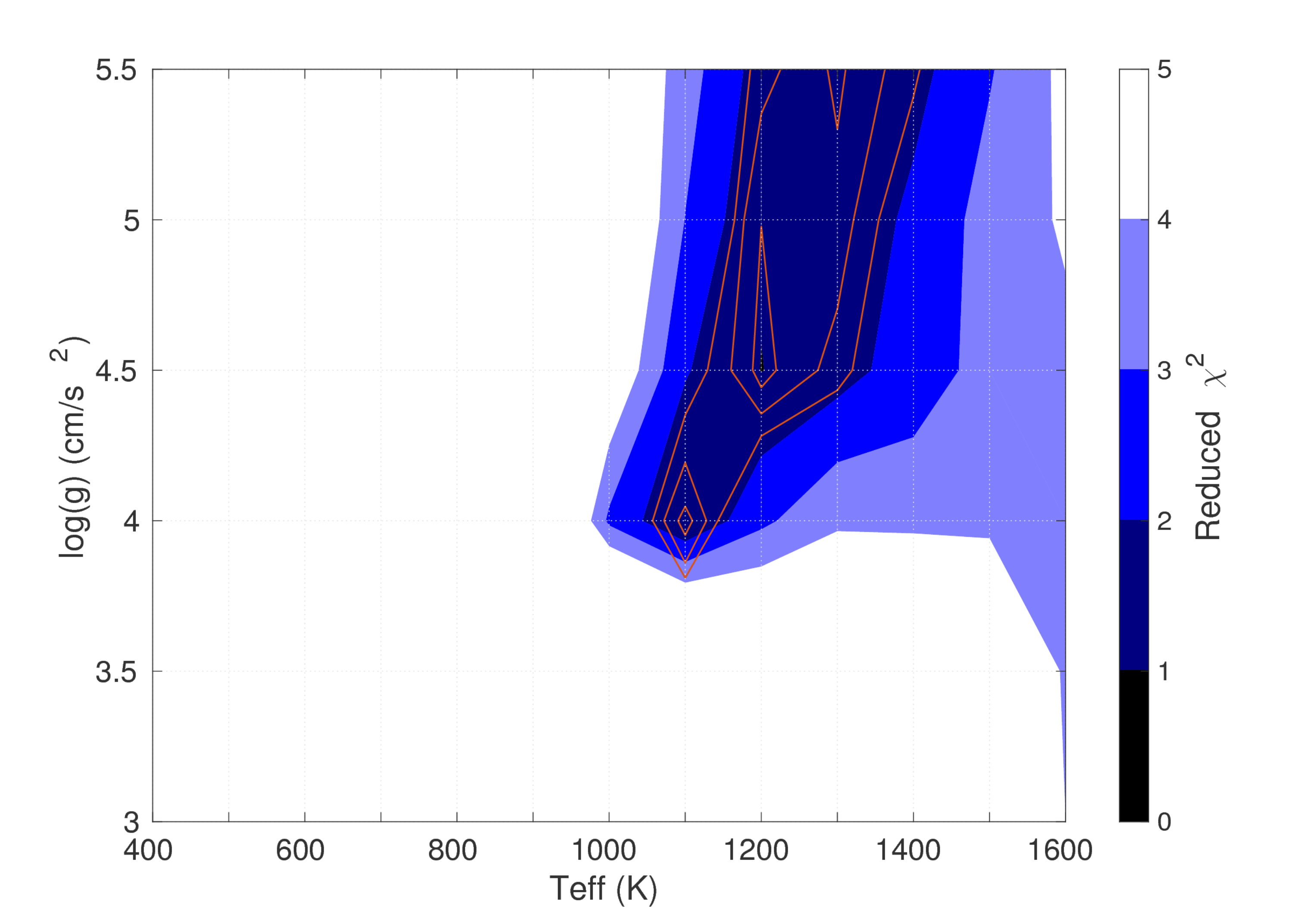}
  \includegraphics[width=\columnwidth]{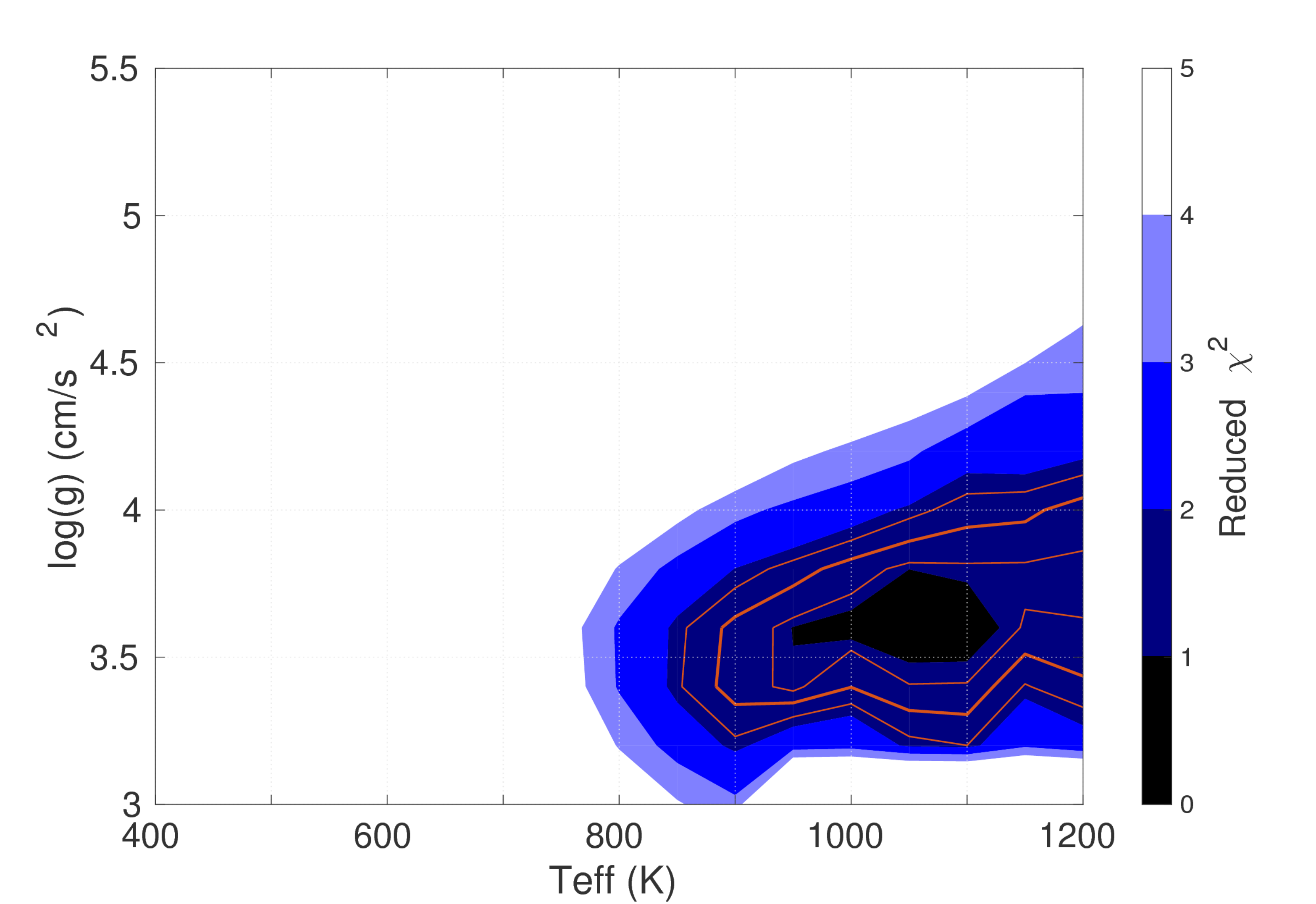}
  \includegraphics[width=\columnwidth]{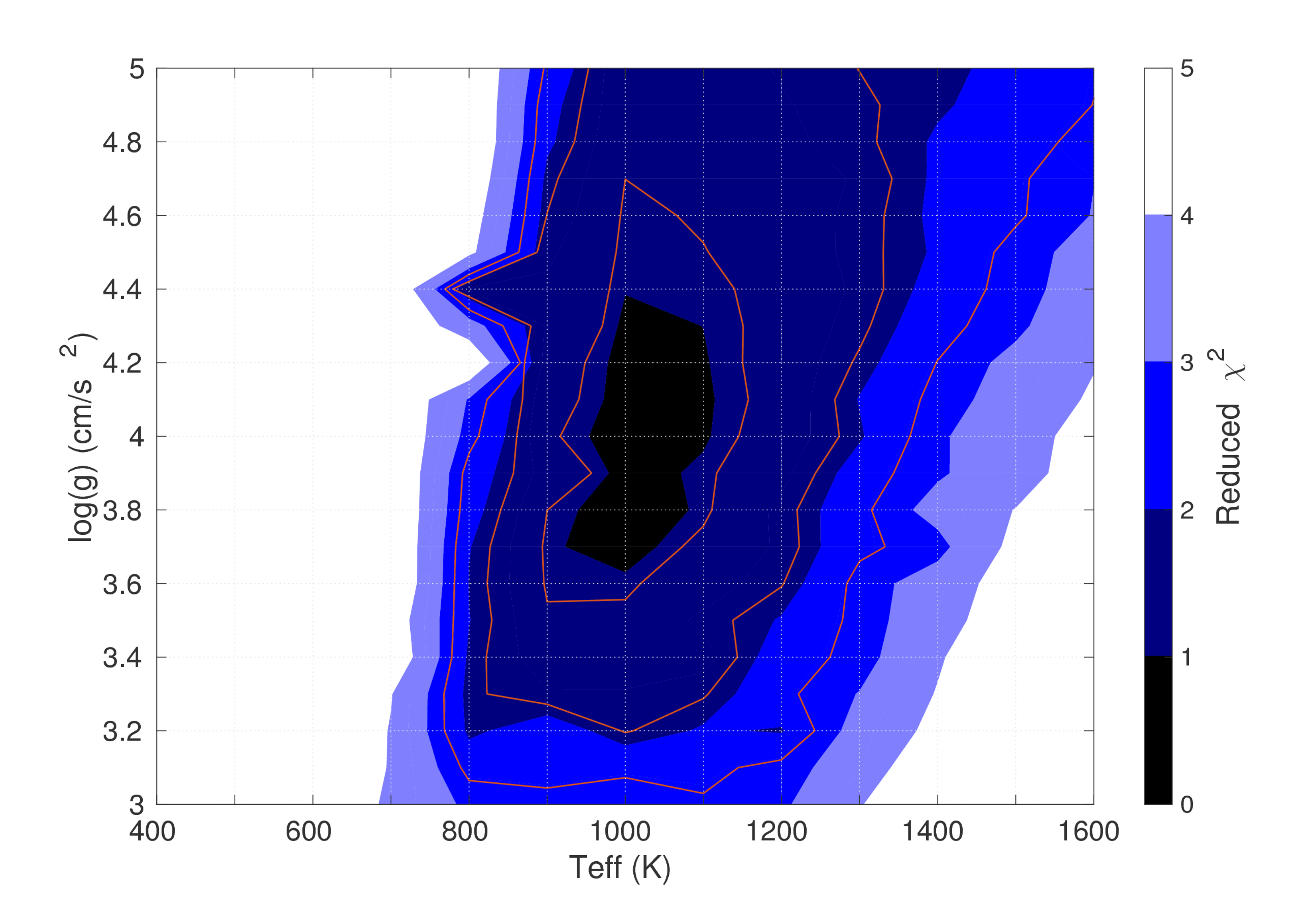}
  \caption{Reduced $\chi^2$ contour map BT-Settl (top), Exo-REM (middle)
      and Exo-REM cloud modified (bottom) models. All the plots shown here
      are for solar metallicities models. The red contours are for 1, 3 and 5
      sigma.}
\label{f:parspace}
\end{figure}

\subsection{Mass limits for other objects in the system}
\label{s:masslimit}

Starting from the contrast limit displayed in Fig.~\ref{f:contrast} and using
the AMES-COND models \citep{2003IAUS..211..325A}, we have obtained the limit in mass
for other possible companions around HR\,2562. This is shown in
Fig.~\ref{f:masslimit}, where the solid lines are obtained assuming an age of
450~Myr. Moreover, with the aim to show the dependency of the limits from the age,
we display in dashed lines the mass limits that are obtained assuming 200 and
750~Myr (these are the lower and upper limits proposed for the age). 
From our findings we should be able to see objects with mass of the order of 10\MJup
at separations larger than 10~au, while at separation larger than 40~au IRDIS
should allow us to detect objects with masses of few \MJup. 

\begin{figure}
  \begin{center}
\centering
\includegraphics[width=0.45\textwidth]{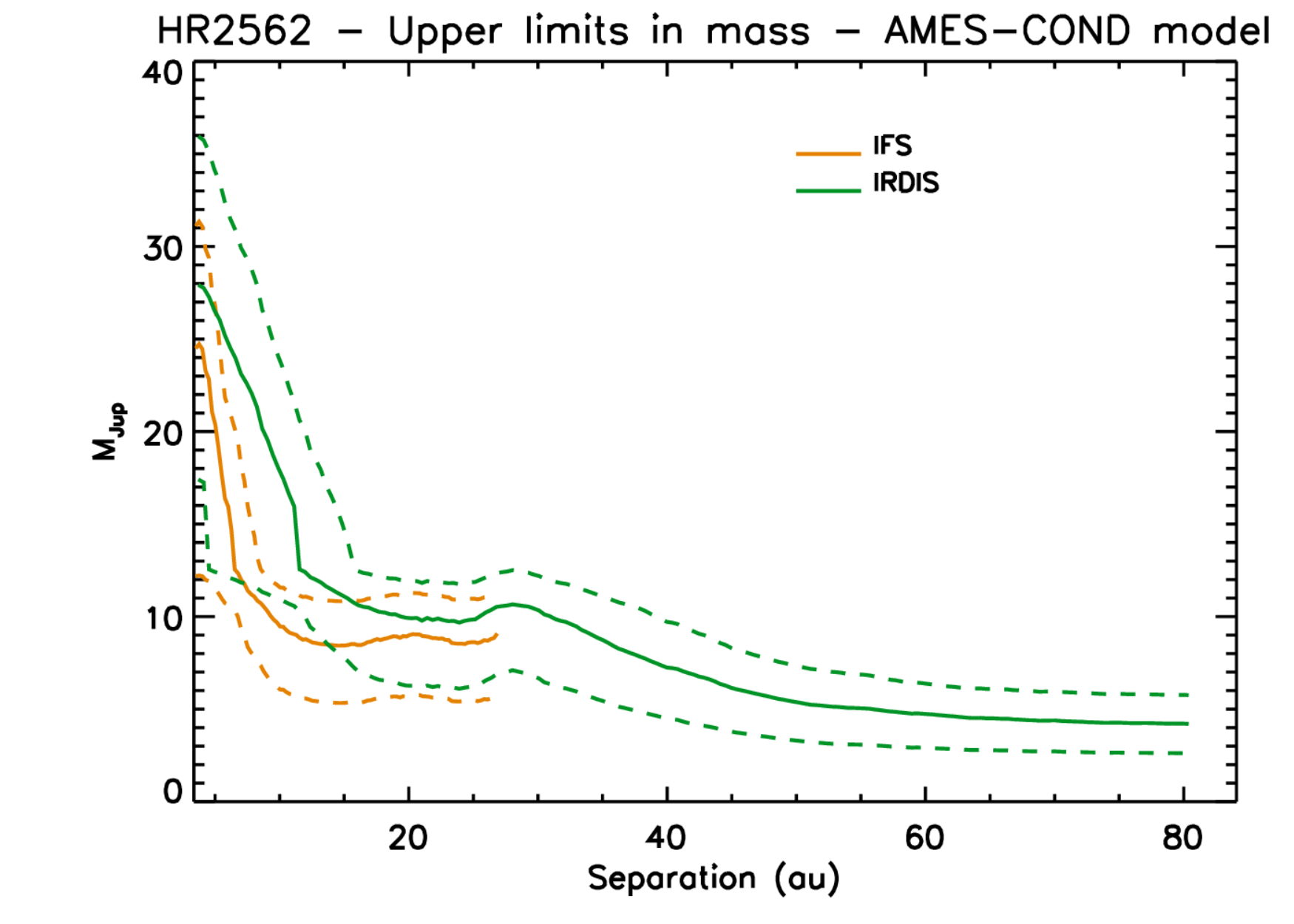}
\caption{Mass limits for possible companions around HR\,2562. The green
  lines are obtained from IRDIS data while orange lines are from IFS data.
  The solid lines are obtained assuming and age of 450~Myr while the dashed
  lines are obtained assuming the lower and upper limits of the age range
  for this star: 200 and 750~Myr.}
\label{f:masslimit}
\end{center}
\end{figure}

%________________________________________________________________________

\section{Conclusions}
\label{conclusion}

In this paper we have presented the high-contrast imaging observations of the star
HR\,2562 obtained with SPHERE. We were
able to recover the low-mass companion previously discovered by \citet{k16}
both with IRDIS and IFS. \par
Using the companion photometry extracted in the Y, J and H band and the
AMES-COND evolutionary models, we derived a mass in the 20-30~\MJup range
and a $\log g\sim$4.7 dex. Conversely, by adopting the AMES-DUSTY models we
obtained
larger masses ($\gtrsim$ 40~\MJup) and $\log g$ of $\approx$ 5.0 dex. \par
The spectro-photometric measurements performed with IFS allowed us
to extract a low-resolution spectrum of the companion in Y and J band,
while IRDIS provided us H broad-band photometry. Fitting the extracted spectrum
with a sample of spectra of low mass objects, we were able to classify
HR\,2562\,B as an early T (T2-T3) spectral type, although the H broad-band
photometry seems more in accordance with late-L spectra.
Our spectral classification is then not in complete accordance with that
by \citet{k16}. The IFS spectrum is very similar to that obtained by GPI
in the J band so that the different spectral classification is probably due
to the different spectral coverage of the two instruments, outside the J band.
In fact, the extension to the Y band allows to SPHERE to see the peak at
1.08~\mic (not visible in GPI), making possible to identify the early-T nature
of the object. On the other hand, the GPI spectrum by \citet{k16} 
extends to the H and K band; their classification was mainly
based on the H and K band observations. Thus, while the presence of
the peak at 1.08~\mic in the IFS spectrum strongly points towards an
early-T type spectrum, several uncertainties
still remain and observations on a wider wavelength range are sorely needed
to completely disentangle the companion spectral classification. \par
It is also possible that some variability affects the spectral
appareance of HR\,2562\,B. Indeed, photometric variability due to
non-uniform cloud coverage has been shown to be more prominent for L/T
transition objects \citep[see e.g. ][]{2014ApJ...782...77B,2015ApJ...799..154M}.
\par
The use of synthetic spectra from atmospheric models allowed us to put
some constraints on the
physical parameters of the companion. Using the BT-Settl models, the
best solution was found for $T_{eff}$=1200$\pm$100~K and $\log{g}$=4.5$\pm$0.5
dex.
The Exo-REM models allowed us to define a range of 950-1200~K for $T_{eff}$
and of 3.4-5.2 for $\log g$, with strong indications for the upper limit of
these intervals. Finally, we have used for our analysis a new version of the
Exo-REM models, which are modified to have a self-consistent treatment of the
clouds.
In this case our findings suggest a $T_{eff}$ in the range 1000-1200~K
and a $\log g$ lower than that found by the other models
(of the order of $\approx$4.0 dex).
However, both the Exo-REM models give strong hints of the presence of clouds
with a high cloud coverage (best fit of 95\%) and non-equilibrium
chemistry. \par
Synthetizing all the results described in the paper, HR\,2562\,B should
have a mass of $32\pm14$~\MJup with $T_{eff}$=$1100\pm200$~K and
$\log g$=$4.75\pm0.41$. 

%----------------------------------------------------------- 

\begin{acknowledgements}
The authors thanks Quinn Konopacky for sharing the GPI spectra of HR\,2562\,B.
We are grateful to the SPHERE team and all the people at Paranal for the great
effort during SPHERE GTO run. \par
This work has made use of the SPHERE Data Center, jointly operated by
OSUG/IPAG (Grenoble), PYTHEAS/LAM/CeSAM (Marseille), OCA/Lagrange (Nice) and
Observatoire de Paris/LESIA (Paris). \par
D.M. acknowledges support from the ESO-Government of Chile Joint Comittee
program 'Direct imaging and characterization of exoplanets'. 
D.M., A.Z., V.D.O., R.G., R.U.C., S.D., C.L. acknowledge support from the ``Progetti Premiali'' funding scheme of the Italian Ministry of Education, University, and Research. J.H. is supported by the French ANR through the GIPSE grant
ANR-14-CE33-0018. This work has been supported by the project PRIN-INAF 2016
The Cradle of Life - GENESIS-SKA (General Conditions in Early
Planetary Systems for the rise of life with SKA). We acknowledge support from
the French National Research Agency (ANR) through the GUEPARD project grant
ANR10-BLANC0504-01. SPHERE is an instrument designed and built by a consortium
consisting of IPAG (Grenoble, France), MPIA (Heidelberg, Germany),
LAM (Marseille, France), LESIA (Paris, France), Laboratoire Lagrange
(Nice, France), INAF-- Osservatorio di Padova (Italy), Observatoire de
Gen\`eve (Switzerland), ETH Zurich (Switzerland), NOVA (Netherlands), ONERA
(France) and ASTRON (Netherlands), in collaboration with ESO. SPHERE was
funded by ESO, with additional contributions from CNRS (France), MPIA
(Germany), INAF (Italy), FINES (Switzerland) and NOVA (Netherlands). SPHERE
also received funding from the European Commission Sixth and Seventh Framework
Programmes as part of the Optical Infrared Coordination Network for Astronomy
(OPTICON) under grant number RII3-Ct-2004-001566 for FP6 (2004-2008), grant
number 226604 for FP7 (2009-2012) and grant number 312430 for FP7 (2013-2016).
\par
This research has benefited from the SpeX Prism Spectral Libraries,
maintained by Adam Burgasser at
http://pono.ucsd.edu/~adam/browndwarfs/spexprism
\end{acknowledgements}

\bibliographystyle{aa}
\bibliography{paper_hip32775_final}

\end{document}